\documentclass{article}
\usepackage{spconf,amsmath,amssymb,graphicx}
\usepackage{enumitem}
\setlist{nosep, leftmargin=14pt}
\usepackage{comment}
\usepackage{hyperref}

\usepackage{mwe} 

\title{Retrieval-Augmented Parkinson's Disease Classification from Resting-State Functional Connectivity}
\name{Tara Niknam,
Sana Hassan Imam\thanks{Sana Hassan Imam and Daniel Kristanto shared senior authorship.},
Daniel Kristanto}

\address{Psychological Methods and Statistics, Department of Psychology,\\
Carl von Ossietzky Universität Oldenburg, 26129 Oldenburg, Germany}
\usepackage{xcolor}

\begin{document}
%
\maketitle
\begin{abstract}
Altered resting-state functional connectivity (FC) is a candidate marker for
the diagnosis of Parkinson's disease (PD). Publicly available PD datasets,
however, are small and fragmented across sites, scanners and clinical
criteria. Transfer learning mitigates data scarcity by reusing a model trained on larger cohorts, but represents their knowledge primarily through a single set of weights shared by all individuals. Motivated by precision neuroimaging, we propose a retrieval-augmented classification (RAC) framework that additionally places each target individual in the
context of similar individuals from external cohorts. A graph convolutional
autoencoder (GAE), trained on external PD and control cohorts, encodes these cohorts
into a neuro-knowledge base. 
Each target individual is encoded using the pretrained GAE, and its nearest neighbors are retrieved from the external knowledge base. Their embeddings are combined with the target embedding using mean aggregation or attention for classification.
A systematic comparison on the TaoWu cohort separated the contributions of
encoder pretraining, retrieval, attention and partial fine-tuning. Pretraining improved mean classification accuracy from 60.5\% to 66.0\%. However, retrieval and attention did not consistently improve performance over the pretrained-only variants. Several retrieval-based configurations nevertheless showed modestly higher ROC-AUC than the randomly initialized baseline. Varying the number of retrieved neighbors and partially fine-tuning the encoder provided no consistent additional gains. Retrieval-neighborhood analysis further showed diagnostically mixed neighborhoods. These findings provide initial evidence that population-learned representations can be complemented by individual-specific retrieval in neuroimaging, although larger datasets are needed to determine when such contextual information improves classification.
\end{abstract}
 
\begin{keywords}
Parkinson's disease, functional connectivity, graph autoencoder, representation learning,
retrieval-augmented classification, transfer learning
\end{keywords}
 
\section{Introduction}
\label{sec:intro}

Parkinson's disease is the fastest-growing neurological disorder~\cite{Rabie2025}, with the number of people living with PD projected to reach 25.2 million by 2050~\cite{Su2025}. Diagnosis remains largely based on clinical motor signs~\cite{Postuma2015}, motivating the search for objective neuroimaging biomarkers. Resting-state functional MRI (rs-fMRI) has revealed reproducible alterations of functional brain organization in PD~\cite{Tahmasian2017}, and FC has increasingly been used for machine-learning-based classification of individuals with PD and healthy controls (HC)~\cite{LopezLimas2025}. 

A major challenge is that PD rs-fMRI cohorts are typically small and distributed across sites, scanners, and clinical populations~\cite{Xu2023,Kristanto2026}. Direct pooling is complicated by acquisition and cohort differences that can substantially affect FC~\cite{Yamashita2019}, as well as regulatory and privacy constraints on sharing individual-level neuroimaging data. Models trained on single cohorts therefore remain limited by small sample sizes and uncertain performance estimates~\cite{Varoquaux2018}. Transfer learning addresses this problem by pretraining models on external datasets before applying them to a target cohort~\cite{Yang2023}. However, the transferred knowledge is primarily encoded in parameters learned across the source population and may not preserve which previously observed individuals are most informative for a particular target individual.

This limitation relates to a central principle of precision neuroimaging: brain organization contains both shared population-level structure and meaningful individual-specific variation~\cite{Gratton2026}. Analogously, transformer models use attention to complement general representations with context-specific information by weighting the relevance of surrounding representations to a given query~\cite{Vaswani2017}. We therefore investigate retrieval-augmented classification, inspired by retrieval-augmented generation~\cite{Lewis2020}, as a way to combine both levels of information. A GAE first learns a common latent representation of FC from external PD and HC. For each target individual, RAC then retrieves the most similar external individuals in this representation space and integrates their embeddings with the representation of the target individual for classification. External cohorts thus serve not only as a source for population-level pretraining, but also as an individual-specific contextual memory. We evaluate whether combining population-learned representations with individual-specific retrieval can provide additional information for PD classification from resting-state FC. 

\newcommand{\Nkb}{169}      
\newcommand{\NkbPD}{139}    
\newcommand{\NkbHC}{30}     
\newcommand{\NtgPD}{20}     
\newcommand{\NtgHC}{20}     
\newcommand{\EpochsRAC}{200}  
\newcommand{\EpochsBase}{100} 
 
\section{Materials and Methods}
\label{sec:methods}

\subsection{Data}
\label{ssec:data}
We used preprocessed resting-state FC matrices of PD and HC cohorts from an
open-access collection of brain network datasets~\cite{Xu2023}. In this
collection, raw MRI data were preprocessed with a common fMRIPrep pipeline,
parcellated with the 100-region Schaefer atlas, and converted into FC
matrices of Pearson correlations between regional BOLD signals. We used three
PD cohorts. The knowledge base comprised $N=\Nkb$ individuals from the
Parkinson's Progression Markers Initiative (PPMI) and Neurocon datasets
($N_{\mathrm{PD}}=\NkbPD$, $N_{\mathrm{HC}}=\NkbHC$); prodromal and SWEDD
individuals of PPMI were excluded. The classification framework was trained and
evaluated exclusively on the TaoWu cohort ($N_{\mathrm{PD}}=\NtgPD$,
$N_{\mathrm{HC}}=\NtgHC$), which is not part of the knowledge base. The
cohorts differ in acquisition and clinical definition: PPMI is a multi-site
study of patients diagnosed within the past two years and not taking PD
medication, whereas TaoWu and Neurocon were acquired on single scanners
and include treated patients~\cite{Xu2023}. Labels were coded as PD $=1$ and
HC $=0$.
 
\subsection{Graph formulation}
\label{ssec:graph}

The signed FC matrices were represented as weighted graphs using absolute
correlations as edge weights, $a_{ij}=|r_{ij}|$, yielding
$\mathbf{A}_s\in[0,1]^{n\times n}$. Thus, positive and negative correlations
of equal magnitude contributed equally. Following the featureless approach
of~\cite{Kipf2017}, we used the identity matrix
$\mathbf{X}=\mathbf{I}_n$ as node features, such that individual-specific
information was encoded by $\mathbf{A}_s$.

\subsection{Unsupervised pretraining of the graph encoder}
\label{ssec:pretrain}

A GAE was trained on the knowledge-base
individuals to reconstruct their FC matrices without using diagnostic labels.
The encoder $f_\theta$ consisted of two graph convolutional
layers~\cite{Kipf2017}:
\begin{equation}
\mathbf{H}^{(l+1)}=
\mathrm{ReLU}\!\left(
\mathbf{A}_s\mathbf{H}^{(l)}\mathbf{W}^{(l)}+\mathbf{b}^{(l)}
\right),
\quad
\mathbf{H}^{(0)}=\mathbf{X},
\label{eq:gcn}
\end{equation}
with $\mathbf{W}^{(0)}\in\mathbb{R}^{100\times64}$,
$\mathbf{W}^{(1)}\in\mathbb{R}^{64\times64}$, and dropout ($p=0.3$) after each
layer. Unlike~\cite{Kipf2017}, $\mathbf{A}_s$ was not degree normalized.
A linear projection mapped the node representations to 128 dimensions,
followed by global mean pooling to obtain the graph-level embedding
$\mathbf{z}_s\in\mathbb{R}^{128}$.

The decoder was an MLP with layers $128\to128\to n^2$, ReLU and dropout
between layers, and a sigmoid output reshaped into the reconstructed adjacency matrix
$\hat{\mathbf{A}}_s\in[0,1]^{n\times n}$, i.e.,\ the absolute FC values that
the encoder received as input. The GAE was trained end-to-end by minimizing the mean squared reconstruction
error
\begin{equation}
\mathcal{L}_{\mathrm{rec}}=
\frac{1}{|\mathcal{B}|\,n^2}
\sum_{s\in\mathcal{B}}
\left\|
\hat{\mathbf{A}}_s-\mathbf{A}_s
\right\|_F^2
\label{eq:rec}
\end{equation}
over mini-batches of 32 graphs using Adam~\cite{Kingma2015} with a learning
rate of $10^{-4}$ and weight decay of $10^{-5}$ for 100 epochs, with gradient
clipping at norm 1. Because the encoder weights are shared across individuals,
the trained encoder can embed unseen graphs and therefore operates
inductively~\cite{Noman2024}. After training, the decoder was discarded and
the encoder retained for subsequent embedding and retrieval.
 
\subsection{Neuro-knowledge base}
\label{ssec:kb}

After pretraining, all \Nkb{} knowledge-base individuals were embedded once
with dropout disabled, and their 128-dimensional embeddings were stored in an
exact Euclidean FAISS index~\cite{Douze2024} for nearest-neighbor retrieval.
 
\subsection{Retrieval-augmented classification}
\label{ssec:rac}

\begin{figure}[t]
\centering
\includegraphics[width=\columnwidth]
{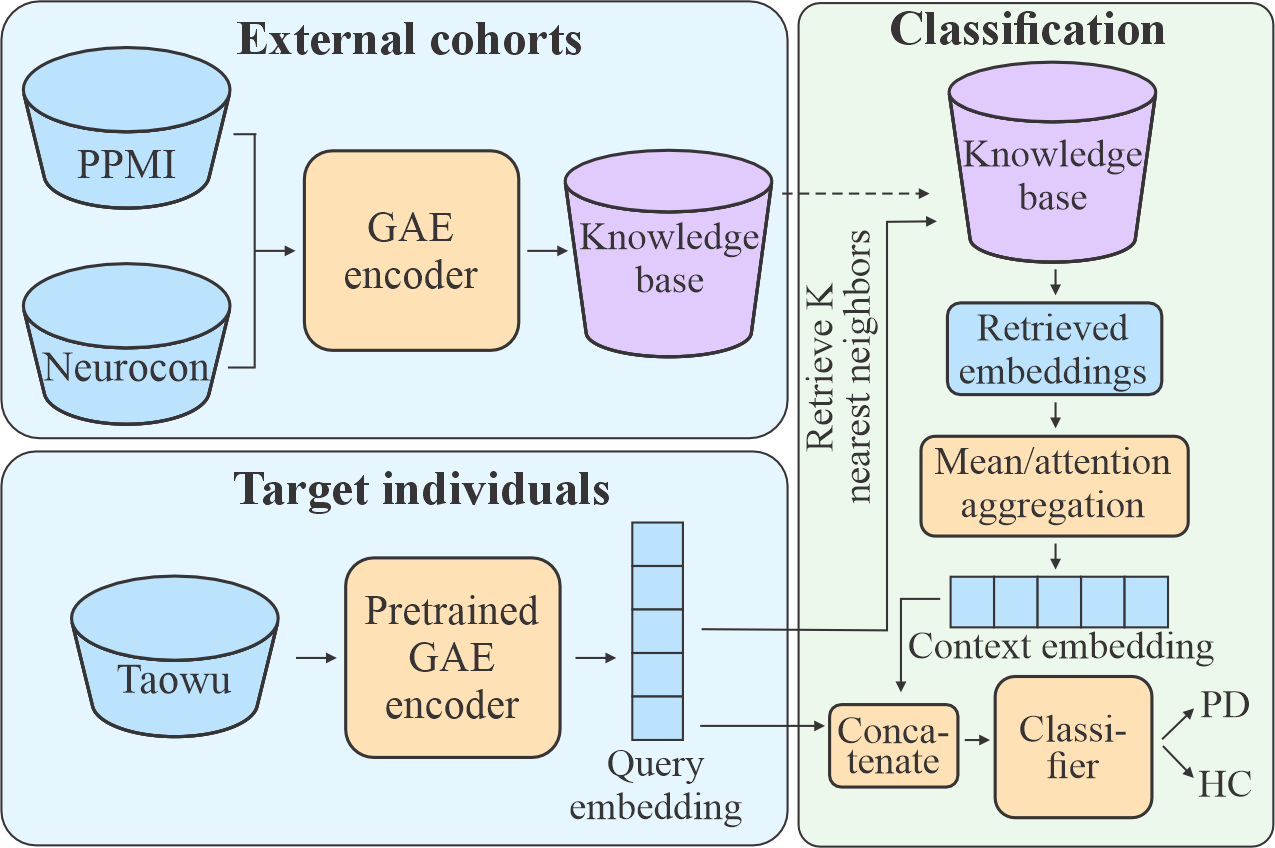}
\caption{Overview of the RAC framework.}
\label{fig:rac_pipeline}
\end{figure}

For each target individual $t$, the (pre-trained or partially fine-tuned)
encoder generated a query embedding
$\mathbf{q}=f_\theta(\mathcal{G}_t)\in\mathbb{R}^{128}$.
The $k\in\{5,10,15\}$ knowledge-base embeddings closest to $\mathbf{q}$ in
Euclidean distance were retrieved,
\[
\mathbf{V}=
[\mathbf{v}_1,\dots,\mathbf{v}_k]
\in\mathbb{R}^{k\times128}.
\]

The retrieved embeddings were combined into a context vector
\begin{equation}
\mathbf{c}=
\sum_{i=1}^{k}\alpha_i\mathbf{v}_i,
\qquad
\alpha_i=
\frac{\exp(s_i)}
{\sum_{j=1}^{k}\exp(s_j)}.
\label{eq:context}
\end{equation}

M2 (Sec.~\ref{ssec:variants}) set $\alpha_i=1/k$, i.e., uniformly averaged
the retrieved embeddings, whereas M3--M5 used attention~\cite{Vaswani2017}
with either parameter-free dot-product or learnable scaled dot-product
scoring. The parameter-free score was
\begin{equation}
s_i=\mathbf{q}^{\top}\mathbf{v}_i.
\label{eq:dot}
\end{equation}
The learnable score was
\begin{equation}
s_i=
\frac{
(\mathbf{W}_Q\mathbf{q}+\mathbf{b}_Q)^{\top}
(\mathbf{W}_K\mathbf{v}_i+\mathbf{b}_K)
}
{\sqrt{d}},
\label{eq:learned}
\end{equation}
with $\mathbf{W}_Q,\mathbf{W}_K\in\mathbb{R}^{d\times d}$ and $d=128$.

The context vector $\mathbf{c}$ was concatenated with the query embedding
$\mathbf{q}$ into $[\mathbf{q};\mathbf{c}]\in\mathbb{R}^{256}$ and passed to
a classification head ($256\to64\to1$, ReLU, dropout $p=0.3$, sigmoid output).
The classifier was trained using binary cross-entropy loss, and, for the
attention variants, attention weights were retained for interpretation.

\subsection{Model variants}
\label{ssec:variants}

We evaluated six configurations. M0 was trained from scratch on TaoWu without
pre-training or retrieval, whereas M1 used the frozen pretrained encoder
without retrieval. M2 added mean retrieval, M3 parameter-free dot-product
attention, and M4 learnable scaled dot-product attention. M5 extended M4 by
fine-tuning only the final encoder projection while keeping both graph
convolutional layers frozen. The external knowledge-base embeddings and FAISS
index remained fixed. Retrieval-based models were evaluated with
$k\in\{5,10,15\}$.

\subsection{Training and evaluation}
\label{ssec:eval}

All configurations were evaluated on TaoWu using repeated stratified five-fold
cross-validation with five repetitions, yielding 25 validation-fold
evaluations per configuration. Identical individual partitions were used
across configurations. Trainable components were reinitialized in each fold;
pretrained models reloaded the external encoder, whereas M0 was initialized
randomly.

Models were trained for up to 100 epochs using Adam
(batch size 16, learning rate $10^{-4}$, weight decay $10^{-4}$)
and binary cross-entropy loss. For M5, the fine-tuned encoder projection used
a learning rate of $10^{-5}$.

A learning-rate scheduler reduced the learning rate based on validation loss,
and early stopping with patience 15 selected the checkpoint with the lowest
validation loss. Retrieval used only the external knowledge base, which
contained no TaoWu individuals. Performance was summarized by mean accuracy
and ROC-AUC across the 25 validation folds, with fold-level variability
reported where applicable. All models were implemented in PyTorch Geometric. Source code is publicly available at \url{https://github.com/tarnik1/practical-project}.

\subsection{Retrieval-neighborhood analysis}
\label{ssec:neighbor_analysis}

Using the frozen pretrained encoder, we retrieved the ten nearest
knowledge-base neighbors for each TaoWu individual within a stratified
five-fold partition and summarized the proportions of PD and HC neighbors
separately for PD and HC target individuals.

\section{Results and Discussion}
\label{sec:Res}

\begin{figure*}[t]
\centering
\includegraphics[width=0.95\textwidth]
{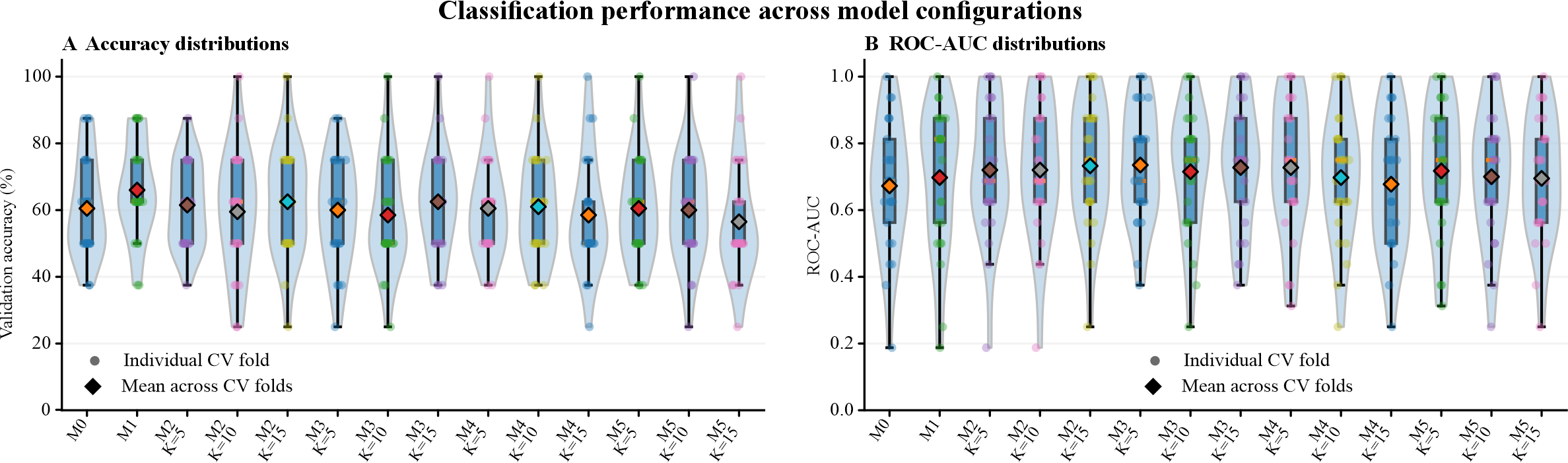}
\caption{Classification performance across model configurations. (A) Validation accuracy and (B) ROC-AUC distributions across 25 fold evaluations per configuration. Dots represent individual fold scores,
and diamonds indicate mean scores. Retrieval-based models (M2--M5) were evaluated with $K\in\{5,10,15\}$ nearest neighbors.}
\label{fig:classification_performance}
\end{figure*}

As shown in Fig.~\ref{fig:classification_performance}, mean validation accuracy improved from M0 to the pretrained-only model M1, but retrieval did not provide a consistent additional improvement across RAC configurations or values of $k$. In contrast, ROC-AUC was modestly higher for several models using the pretrained representation and retrieval than for the randomly initialized baseline, although the distributions across cross-validation folds overlapped substantially. Increasing the number of retrieved neighbors from $k=5$ to $k=15$ produced no consistent improvement in either metric. These results suggest that population-level pretraining and retrieval may influence the ranking of PD and HC individuals, although retrieval did not yield a consistent additional gain in threshold-dependent accuracy.

The diagnostic composition of the retrieved neighborhoods is summarized in Table~\ref{tab:neighbor_composition}. PD target individuals retrieved on average $79.5\pm13.6\%$ PD and $20.5\pm13.6\%$ HC neighbors, whereas HC target individuals retrieved $81.0\pm12.9\%$ PD and $19.0\pm12.9\%$ HC neighbors. Thus, the neighborhoods were diagnostically mixed and showed similar compositions for PD and HC target individuals. This indicates that similarity in the unsupervised FC representation does not simply reproduce diagnostic group membership. At the same time, the predominance of PD individuals in the external knowledge base likely contributes to the observed neighborhood composition. More broadly, this result is consistent with the motivation underlying RAC and with the perspective of precision neuroimaging, which emphasizes that meaningful individual variation may not align directly with categorical group boundaries. The retrieval framework operationalizes this idea by identifying the external individuals most similar to each target individual rather than relying exclusively on population-level representations.

\begin{table}[t]
\centering
\caption{Diagnostic composition of the ten nearest knowledge-base neighbors ($k=10$) using the frozen pretrained encoder. Values are mean $\pm$ SD.}
\vspace{0.7cm}
\label{tab:neighbor_composition}
\footnotesize
\setlength{\tabcolsep}{3pt}
\begin{tabular}{@{}lccc@{}}
\hline
Target & $n$ & PD neighbors (\%) & HC neighbors (\%) \\
\hline
PD & 20 & $79.5\pm13.6$ & $20.5\pm13.6$ \\
HC & 20 & $81.0\pm12.9$ & $19.0\pm12.9$ \\
\hline
\end{tabular}
\end{table}

\section{Conclusion}
\label{sec:Con}

This study investigated retrieval-augmented classification as a framework for combining population-level representation learning with individual-specific contextual information in PD functional connectivity. Although retrieval-based RAC configurations did not consistently improve accuracy beyond the pretrained-only model, several configurations showed modestly higher ROC-AUC, while the retrieved neighborhoods revealed that FC similarity was not restricted to individuals sharing the same diagnosis. Together, these findings suggest that external cohorts can provide information beyond a single population-trained representation, but that the present retrieval strategy does not yet translate this information into consistently improved classification.

The framework was motivated in part by transformer and retrieval-based approaches in large language models, where attention and retrieval benefit from very large training corpora and rich representation spaces. The present results provide initial proof-of-concept evidence that a related principle can be implemented in neuroimaging, by combining a population-learned representation with information retrieved specifically for each target individual. However, the substantially smaller scale of available neuroimaging datasets likely limits the extent to which such individualized retrieval can be learned reliably. From a precision neuroimaging perspective, the framework offers a computational way to complement population-level learning with individual-specific contextualization, although the present results indicate that larger and more diverse datasets are needed to determine when such personalization improves clinically relevant prediction.

A key next step is therefore to evaluate the framework in substantially larger and more diverse datasets. Larger cohorts would allow individual-specific neighborhoods to be estimated more reliably, provide richer variability for representation learning, and enable a more robust assessment of whether retrieval contributes information beyond population-level pretraining. They may also help determine whether retrieval captures clinically meaningful heterogeneity beyond binary PD versus HC labels. Future extensions could further enrich the knowledge base with clinical metadata, such as diagnosis, medication status, or other individual-level characteristics, allowing retrieval to incorporate not only imaging similarity but also clinically relevant contextual information.

\section{Acknowledgments}
\label{sec:acknowledgments}

This work was supported by the Young Researchers Fellowship from the University
of Oldenburg and by a grant from the German Research Foundation (DFG) to Andrea Hildebrandt, Carsten Gießing, Christiane Thiel, and Stefan Debener as part of the DFG priority program "META-REP: A Meta-scientific Program to Analyze and Optimize Replicability in the Behavioral, Social, and Cognitive Sciences" (SPP 2317; Project number 464552782).

Neurocon data used in this article were obtained from the NEUROCON project
(84/2012), financed by UEFISCDI.

PPMI data used in this article were obtained from the PPMI database
\url{https://www.ppmi-info.org/}. PPMI -- a
public-private partnership -- is funded by The Michael J. Fox Foundation for
Parkinson's Research and funding partners, including 4D Pharma, AbbVie Inc.,
AcureX Therapeutics, Allergan, Amathus Therapeutics, Aligning Science Across
Parkinson's (ASAP), Avid Radiopharmaceuticals, Bial Biotech, Biogen,
BioLegend, Bristol Myers Squibb, Calico Life Sciences LLC, Celgene Corporation,
DaCapo Brainscience, Denali Therapeutics, The Edmond J. Safra Foundation,
Eli Lilly and Company, GE Healthcare, GlaxoSmithKline, Golub Capital,
Handl Therapeutics, Insitro, Janssen Pharmaceuticals, Lundbeck, Merck \& Co.,
Inc., Meso Scale Diagnostics, LLC, Neurocrine Biosciences, Pfizer Inc.,
Piramal Imaging, Prevail Therapeutics, F. Hoffmann-La Roche Ltd and its
affiliated company Genentech Inc., Sanofi Genzyme, Servier, Takeda
Pharmaceutical Company, Teva Neuroscience, Inc., UCB, Vanqua Bio, Verily Life
Sciences, Voyager Therapeutics, Inc., and Yumanity Therapeutics, Inc.
For up-to-date information on the study, visit www.ppmi-info.org.

The authors have no relevant financial or non-financial interests to disclose.

\bibliographystyle{IEEEbib}
\bibliography{strings,refs}

 \vspace{2.0cm}
 \vspace{1.5cm}
 \vspace{1.5cm}

\end{document}